\documentstyle[camera,prbplug,psfig]{article}
 
\begin{document}

\title{
Orbital-dependent two-band superconductivity
in MgB$_2$
}

\author{Takashi Yanagisawa$^{1,2}$ and Hajime Shibata$^1$ 
}

\address{$^1$Nanoelectronics Research Institute, National Institute of 
Advanced Industrial Science and Technology (AIST) Tsukuba Central 2, 1-1-1 Umezono, 
Tsukuba, Ibaraki 305-8568, Japan\\
$^2$Graduate School of Science, Osaka University, Toyonaka 560-0043, Japan
}

\maketitle

\begin{abstract}
We show that a two-band model with ${\bf k}$-dependent superconducting gaps
well describes
the transmission and optical conductivity measured for MgB$_2$ thin films.
It is also shown that the two-band anisotropic model consistently describes the
specific-heat jump and thermodynamic critical magnetic field $H_c$.
A single-gap anisotropic model is shown to be insufficient to understand consistently
optical and thermodynamic behaviors.
In our model, the pairing symmetry in each band has an anisotropic characteristic 
which is
determined almost uniquely;
the superconducting gap in the $\sigma$-band has anisotropy in the ab-plane and
the gap in the $\pi$-band has a prolate form exhibiting anisotropy in the
c-direction.  The anisotropy in the $\sigma$-band produces rather small effects on the 
physical properties compared to the anisotropy in the $\pi$-band.
\end{abstract}


Recently discovered superconductor MgB$_2$ with a relatively high T$_c$=39K
has attracted much attention in condensed-matter physics\cite{nag01}.
An $s$-wave superconductivity (SC) was soon established by experiments, e.g., a
coherence peak in $^{11}$B nuclear relaxation rate\cite{kot01} and its exponential
dependence at low temperatures\cite{yan01,man02}.
An isotope effect has suggested phonon-mediated $s$-wave superconductivity\cite{bud01}.
In contrast to its standard properties, there have been several reports indicating
unusual properties of the superconductivity of MgB$_2$.
Several studies reported two different superconducting gaps:
a gap much smaller than the expected BCS value and that comparable
to the BCS value given by $2\Delta=3.53k_BT_c$.
Their ratio is estimated to be $\Delta_{min}/\Delta_{max}\sim 0.3-0.4$ using several
experiments.\cite{yan01,tsu01,sza01,che01,giu01,bou01}
It is also reported that the specific-heat jump and the critical magnetic field
are reduced compared to the $s$-wave BCS theory.\cite{yan01,bou01}
The other typical characteristic is a strongly anisotropic upper critical field in 
$c$-axis-oriented MgB$_2$ films and single crystals of 
MgB$_2$.\cite{lim01,xu01,ang02}

The unusual properties of MgB$_2$ suggest an anisotropic $s$-wave superconductivity
or a two-band superconductivity.  
The band structure calculations predicted multibands originating from
$\sigma (2p_{x,y})$ and $\pi (2p_z)$ bands.\cite{kor01}
In the ARPES measurements performed in single crystals of MgB$_2$ three distinct
dispersions approaching the Fermi energy were reported.\cite{uch02}

There have been several studies on the anisotropy of a superconducting 
gap\cite{haa01,mir02,nak02,tew02,bou02}. 
The two-gap model is shown to consistently describe the specific heat\cite{nak02,bou02}
and the upper critical field $H_{c2}$\cite{mir02} with the adoption of the effective 
mass approach.
In this paper, we examine optical properties and thermodynamics to determine the 
{\bf k}-dependence of the gaps.
The ${\bf k}$-dependence of two gaps in MgB$_2$ has never been investigated thus far.
We show that the optical transmittance, conductivity, specific-heat jump,
and thermodynamic critical field $H_c$ are well described by a two-band 
superconductor model with different anisotropies in ${\bf k}$-space.
The symmetry in ${\bf k}$-space is determined almost uniquely according to these
experiments.
First, we show that the single-gap model is insufficient to understand consistently
optical and thermodynamic behaviors.  Second, the two-gap model with different
symmetries in ${\bf k}$-space is shown to be sufficient to understand optical and
thermodynamic properties.


The optical conductivity for anisotropic s-wave SC is
investigated and compared with available data for MgB$_2$.
A simple angle-dependent generalization of the Mattis-Bardeen formula\cite{mat58} 
is used to calculate the optical conductivity.  
The density of states $N(\epsilon)=\epsilon/\sqrt{\epsilon^2-\Delta^2}$ is
generalized to
$N(\epsilon)=\langle{\rm Re}\epsilon/\sqrt{\epsilon^2-\Delta_{{\bf k}}^2}\rangle_k$,
where the bracket indicates the average over the Fermi surface.
We employed the following formula at T=0:
\begin{eqnarray}
\frac{\sigma_{1s}(\omega)}{\sigma_{1n}}&&
=\frac{1}{\omega} \int_0^{\omega}dE [ N(E)N(\omega-E)\nonumber\\
&&- \langle {\rm Re} \frac{\Delta_{\bf k}}{\sqrt{E^2-\Delta_{\bf k}^2}} \rangle _k
\langle {\rm Re} \frac{\Delta_{\bf k'}}{\sqrt{(\omega-E)^2-\Delta_{{\bf k'}}^2}}
\rangle _{k'} ] .\nonumber\\
\end{eqnarray}
$\sigma_{2s}/\sigma_{1n}$ is similarly generalized.
The anisotropic order parameters considered in this paper are:
\begin{eqnarray}
\Delta_{c1}({\bf k})&=& \Delta (1+a{\rm cos}(2\theta)),\\
\Delta_{c2}({\bf k})&=& \Delta (1-b{\rm cos}^2(\theta)),\\
\Delta_{ab}({\bf k})&=& \Delta (1+c{\rm cos}(6\phi)).
\end{eqnarray}
Here, $\theta$ and $\phi$ are the angles in the polar coordinate where $\theta$ is 
the polar angle with respect to the c-axis.  The parameters $a$, $b$ and $c$ 
determine the anisotropy.  
$\Delta_{c1}$ is a prolate form gap for $a>0$ and
is oblate for $a<0$.
$\Delta_{c2}$ ($b>0$) shows the same anisotropy as $\Delta_{c1}$ for $a<0$. 
$\Delta_{ab}$ indicates a cylindrical gap with anisotropy in the a-b plane
corresponding to the $\sigma$ band.
The integral in eq.(1) is detemined numerically by writing the average over the
Fermi surface with elliptic functions.

\begin{figure}
\centerline{\psfig{figure=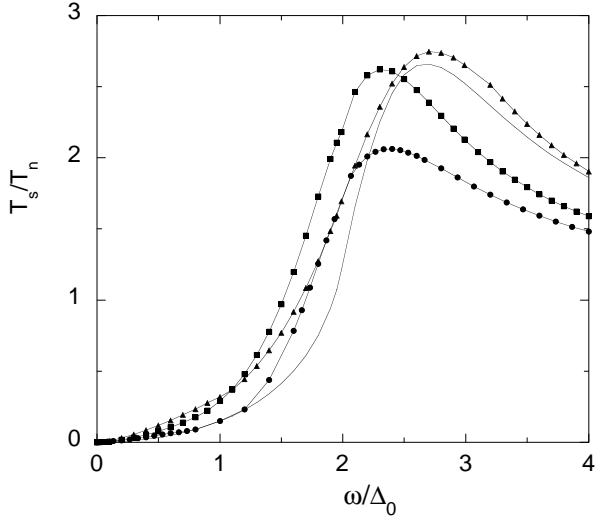,width=8cm}}
\caption{
Transmission $T_S/T_N$ for anisotropic $s$-wave models.
The solid curve without marks is the Mattis-Bardeen result.
Squares, triangles and circles are for the prolate, ab-plane anisotropic, and oblate
gaps, respectively.  The oblate form shows a small increase compared to other types.
For the oblate gap, $\Delta_0=\Delta_{max}$, and for other types, $\Delta_0=\Delta$.
}
\label{fig2}
\end{figure}

\begin{figure}
\centerline{\psfig{figure=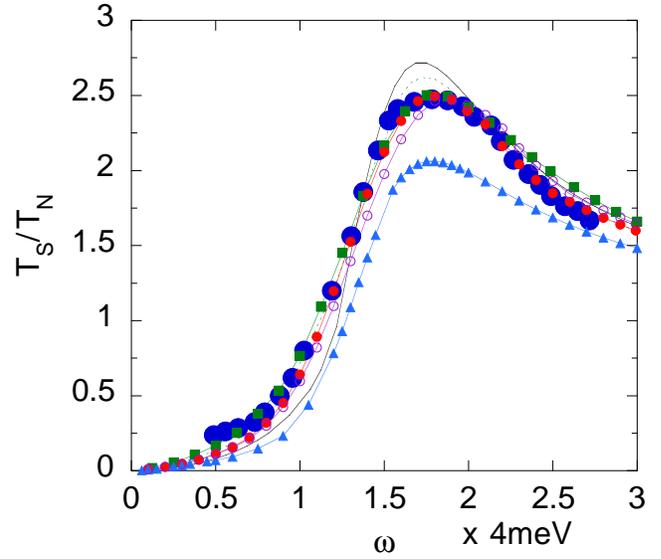,width=\columnwidth}}
\caption{
Transmission $T_S/T_N$ for the two-band model.
The data points (large solid circles) are taken from ref.\protect\CITE{kai02} at 
$T=4$K.
The solid line without marks shows the Mattis-Bardeen result with $2\Delta=5$meV.
The dotted line is for the single gap of prolate type with $2\Delta_{max}=9$meV and 
$a=0.5$.  Triangles are for the single-gap model of oblate form with
$2\Delta_{max}=9$meV and $a=0.33$.
Others are for the two-band gap model where the ab-plane anisotropic form for
the $\sigma$-band
and the prolate form for the $\pi$-band are assumed.  The parameters are the following.
Solid circles: $2\Delta_{max}=8.5$meV and $c=0.33$ ($\sigma$-band: weight 0.45);
$2\Delta_{max}=6$meV and $a=0.33$ ($\pi$-band: weight 0.55).
Open circles: $2\Delta_{max}=9$meV and $c=0.33$ ($\sigma$-band: weight 0.45);
$2\Delta_{max}=6$meV and $a=0.33$ ($\pi$-band: weight 0.55).
Squares: $2\Delta_{max}=10$meV and $c=0.5$ ($\sigma$-band: weight 0.4);
$2\Delta_{max}=7.5$meV and $a=0.5$ ($\pi$-band: weight 0.6).
}
\label{fig3}
\end{figure}

First, we show that the one-band model is insufficient to understand consistently 
optical and thermodynamic behaviors.
In Fig.\ref{fig2} the transmission $T_{S}$ at
$T=0$ is shown as a function of the frequency $\omega$.
The following phenomenological expression 
for $T_S/T_N$
is employed to determine the transmission curve theoretically,\cite{yana01,glo57}
\begin{equation}
\frac{T_S}{T_N}= \frac{1}{[T_N^{1/2}+(1-T_N^{1/2})(\sigma_1/\sigma_n)]^2+
[(1-T_N^{1/2})(\sigma_2/\sigma_n)]^2},
\end{equation}
where $\sigma_1$ and $\sigma_2$ are real and imaginary parts of the optical
conductivity, respectively.
$T_N$ is determined from the
expression for the ratio of the power transmitted with a film to that transmitted
without a film given as
$T_N= 1/[1+\sigma_n d Z_0/(n+1)]^2$.
Here, $d$ is the film thickness, $n$ is the index of refraction of the 
substrate,
and $Z_0$ is the impedance of free space.  
We have assigned the following
values:  $d= 10^{-6}$ cm, $n\approx 3$, $Z_0=377 \Omega$, and
$\sigma_n\approx 8\times 10^3 \Omega^{-1}$cm$^{-1}$.
Then we obtain $T_N\simeq 0.014$.
The theoretical curves for $T_S/T_N$ are shown in Fig.\ref{fig2}; they have peaks
near $\omega\sim 2\Delta_0$.
For the oblate, its peak shows an increase only twice the normal state
value, while the prolate and ab-plane anisotropies show more than
twofold increases.   The experiments show an approximately 2.5-fold 
increase\cite{kai02}
which supports the prolate or ab-plane anisotropic symmetry.  
However, the
temperature dependence of the ratio $H_{c2}^{ab}/H_{c2}^c$, which increases as the
temperature decreases\cite{ang02}, indicates that $\Delta({\bf k})$
has an oblate form instead of a prolate form\cite{haa01} in contrast to $T_S/T_N$.
It is also difficult to describe the thermodynamic quantities such as the
specific-heat jump at $T=T_c$ and the thermodynamic critical magnetic field $H_c$
within the single-gap model consistently.
The specific-heat jump at $T_c$ is given by
\begin{equation}
\frac{\Delta C(T_c)}{\gamma_C T_c}=\frac{12}{7\zeta (3)}
\frac{\langle z^2\rangle ^2}{\langle z^4\rangle },
\end{equation}
where $\gamma_C$ is the specific-heat coefficient and $z$ is an anisotropic factor
of the gap function.  $\langle z^n\rangle$ is the average of $z^n$ over the
Fermi surface.
The experiments indicate that this value is in the range of 0.76$\sim$ 
0.92\cite{yan01,bou01}; the fitting parameters must be $a\sim 0.3$,
$-a\sim 0.5$ and $c\sim 0.3$ for the prolate, oblate and ab-anisotropic types, 
respectively.
We must assign different values to parameters $a$ and $c$ in order to explain
the thermodynamic critical magnetic field $H_c$.
The ratio of $H_c(T=0)$ to the BCS value is given as 
\begin{equation}
H_c(T=0)^2/\gamma_CT_c^2=(6\pi/{\rm e}^{2\gamma})\langle z^2\rangle=
5.94\langle z^2\rangle.  
\end{equation}
Thus to be consistent with the experimental results\cite{bou01},
$\langle z^2\rangle$ should be less than 1;
$a$ should be small, $a\sim 0.07$, for the prolate form, and the ab-plane anisotropic
and oblate forms $(a<0$) are ruled out since $\langle z^2\rangle >1$.
In Table I, we summarize the status for the single-gap anisotropic $s$-wave model 
applied to MgB$_2$. As shown here, it is difficult to understand the physical behaviors
measured using several experimental methods consistently within the single-gap model.

\begin{figure}
\centerline{\psfig{figure=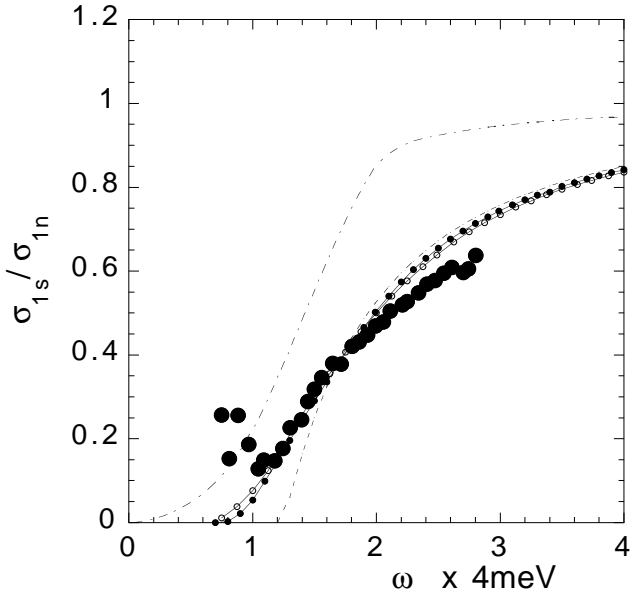,width=\columnwidth}}
\caption{
Real part of the optical conductivity for the two-band anisotropic model.
The data points (large solid circles) are taken from ref.\protect\CITE{kai02}
The parameters for solid circles are
$2\Delta_{max}=8.5$meV and $c=0.33$ ($\sigma$-band: weight 0.45);
$2\Delta_{max}=6$meV and $a=0.33$ ($\pi$-band: weight 0.55).
The parameters for open circles are
$2\Delta_{max}=10$meV and $c=0.5$ ($\sigma$-band: weight 0.4);
$2\Delta_{max}=7.5$meV and $a=0.5$ ($\pi$-band: weight 0.6).
The dashed line indicates the results obtained using the Mattis-Bardeen formula 
with $2\Delta=5$meV.
The dash-dotted line denotes the conductivity for the $d$-wave 
gap.\protect\cite{yana01}
}
\label{fig4}
\end{figure}

\begin{figure}
\centerline{\psfig{figure=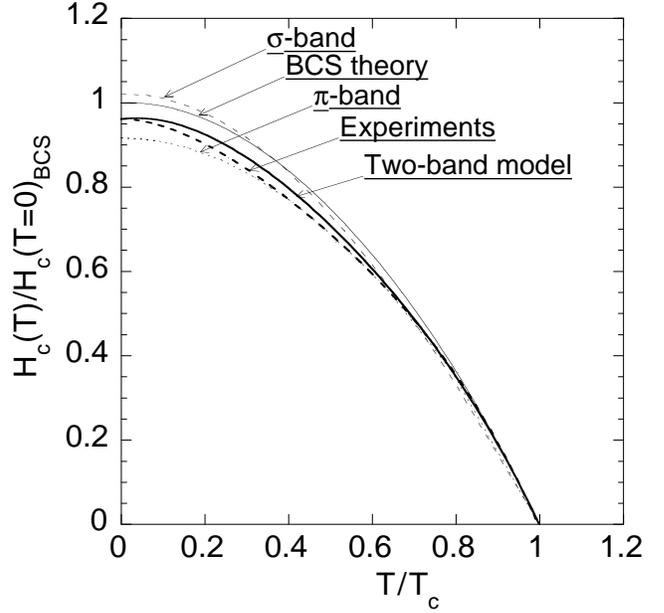,width=\columnwidth}}
\caption{
Thermodynamic critical magnetic field $H_c(T)$ normalized by $H_c(T=0)_{BCS}$.
The bold dashed curve indicates data from ref.\protect\CITE{bou01} and the bold solid 
curve indicates those obtained using the
present two-band anisotropic model.  The thin solid curve indicates the BCS results.
The results for the $\sigma$-band (ab-plane anisotropic) and the $\pi$-band 
(prolate form) are also shown.
}
\label{fig5}
\end{figure}

Here, a two-band model with two different anisotropies is investigated.
We assume that the hybridization between $\sigma$ and $\pi$ bands is negligible that
the optical conductivity is given by
\begin{equation}
\sigma= w_{\sigma}\sigma_{\sigma}+w_{\pi}\sigma_{\pi},
\end{equation}
where $\sigma_{\sigma}$ and $\sigma_{\pi}$ denote the contributions from $\sigma$-
and $\pi$-bands, respectively.
The transmission $T_S/T_N$ in Fig.\ref{fig3} shows that the theoretical
curve is in good agreement with the experimental curve.
The effect of $ab$-anisotropy for the transmission $T_S/T_N$ is small.
The optical conductivity is also described well by the two-band model as shown
in Fig.\ref{fig4}.
We assign the following parameters to the best fit model in Figs.\ref{fig3} and 
\ref{fig4}.
The $\sigma$-band has ab-plane anisotropy with $c\approx$ or less than 0.33 and
the $\pi$-band has the prolate form gap (cigar type) with $a\approx 0.33$.
The ratio of the weight of the $\sigma$-band to that of the $\pi$-band is 0.45/0.55, 
which
agrees with penetration depth\cite{man02} and band structure calculations\cite{bel01}.
The ratio of the minimum gap to the and maximum gap is 0.35, which is in the range of
previously reported experimental values.\cite{tsu01,giu01}
Let us mention here that the two-band isotropic model ($z=1$) describes inconsistently
the observed $\sigma_{1s}/\sigma_{1n}$ since a shoulder-like structure is
predicted in the two-gap isotropic model if the two gaps have different magnitudes.
In Fig.\ref{fig5} the thermodynamic critical magnetic field $H_c(T)$ is shown for the
single-band and two-band models with available data.\cite{bou01}
We have simply assumed that the total free energy is given by the sum of two
contributions from $\sigma$- and $\pi$-bands: 
$\Omega=w_{\sigma}\Omega_{\sigma}+w_{\pi}\Omega_{\pi}$.
The experimental behavior is well explained by the two-band anisotropic model
using the same parameters as those for $T_S/T_N$ and $\sigma_{1s}/\sigma_{1n}$.
We show several characteristic values obtained from the two-band model
in Table II.  Results of
analyses of $H_{c2}$ and specific heat using the effective mass approach are
consistent with those obtained using the two-band model.\cite{mir02,nak02,bou02}
It has been reported that the increasing nature of $H_{c2}^{ab}/H_{c2}^c$ with
decreasing temperature is explained by the two-Fermi surface model.\cite{mir02}
The specific-heat coefficient $\gamma$ in magnetic fields seems consistent
with that of the multiband superconductor.\cite{nak02,bou02}

We have examined the transmittance,  optical conductivity,
specific-heat jump and thermodynamic critical magnetic field $H_c$ of MgB$_2$ based 
on the two-band 
anisotropic s-wave pairing model.
We have shown that the two-gap model with different symmetries in ${\bf k}$-space can 
explain 
the experimental results consistently.

\onecolumn
\begin{table}
\caption{Anisotropic parameters in the SC gap function used to fit several physical 
quantities.  The upper four rows are for the single-SC gap model and the last row is 
for
the two-band anisotropic model for comparison.
The cross indicates that we cannot fit experimental data by the corresponding
$z$ factor.  $\Delta$ in the column $H_{c2}$ indicates that
experiments are explained qualitatively but not quantitatively.\protect\cite{haa01}
The big circle show that we can fit the data using the same parameters
in the column $\sigma_1$.
For the two-gap model the
anisotropy parameter in the $\sigma$-band can be small since the $\sigma$-band
anisotropy produces only small effects on physical quantities.
}
\label{table1}
\begin{tabular}{cccccccc} 
\hline
 & $z$ & $\sigma_1$ & $T_S/T_N$ & $\Delta C$ & $H_c(0)$ &
$H_{c2}$  \\
\hline
Cigar-type & $1+a{\rm cos}(2\theta)$ & $a\sim 0.5$ & $\sim 0.3$ & $\sim 0.3$ 
& $\sim 0.07$      & {\large $\times$} \\
Pancake    & $1-a'{\rm cos}(2\theta)$& $a'\sim 0.6$& {\large $\times$} & $\sim 0.5$  
& {\large $\times$} & $\Delta$  \\
Pancake    & $1-b{\rm cos}^2(\theta)$& $b\sim 0.75$& {\large $\times$} & $\sim 0.66$
& $\sim 0.08$      & $\Delta$  \\
In-plane & $1+c{\rm cos}(6\phi)$     & $c\sim 0.5$ & $\sim 0.3$ & $\sim 0.3$
& {\large $\times$} &   \\
Two-band & ($\sigma$ band) & $c\leq 0.3$ & $\bigcirc$ & $\bigcirc$ & $\bigcirc$ &  \\ 
         & ($\pi$ band)    & $a\sim 0.3$ &  & &  & \\ 
\hline
\end{tabular}
\end{table}

\begin{table}
\caption{Several physical quantities obtained by the two-band model
with $c\sim 0.33$ ($\sigma$ band) and $a\sim 0.33$ ($\pi$ band).
}
\label{table2}
\begin{tabular}{ccccc} \hline
 & $w_{\sigma}/w_{\pi}$ & $\Delta_{min}/\Delta_{max}$ & 
$\frac{\Delta C(T_c)}{\Delta C(T_c)_{BCS}}$  & $\frac{H_c(0)}{H_c(0)_{BCS}}$   \\
\hline
Two-band & 0.45/0.55 & $\sim 0.35$ & $\sim 0.82$ & $\sim 0.95$  \\ 
Exp.     & 0.45/0.55  & $0.3-0.4$ & $0.76-0.92$
& 0.96 \\
\hline
\end{tabular}
\end{table}
\twocolumn


\begin{thebibliography}{9}
\bibitem{nag01}J. Nagamatsu, N. Nakagawa, T. Muranaka, Y. Zenitani and J. Akimitsu: 
Nature {\bf 410} (2001) 63.
\bibitem{kot01}H. Kotegawa, K. Ishida, Y. Kitaoka, T. Muranaka and J. Akimitsu: 
Phys. Rev. Lett. {\bf 87} (2001) 127001.
\bibitem{yan01}H. D. Yang, J. Y. Lin, H. H. Li, F. H. Hsu, C. J. Liu, S. C. Li, 
R. C. Yu
and C. Q. Jin: Phys. Rev. Lett. {\bf 87} (2001) 167003.
\bibitem{man02}F. Manzano, A. Carrington, N. E. Hussey, S. Lee, A. Yamamoto and
S. Tajima: Phys. Rev. Lett. {\bf 88} (2002) 047002.
\bibitem{bud01}S. L. Bud'ko, G. Lapertot, C. Petrovic, C. E. Cunningham, N. Anderson
and P. C. Canfield: Phys. Rev. Lett. {\bf 86} (2001) 1877.
\bibitem{tsu01}S. Tsuda, T. Yokoya, T. Kiss, Y. Takano, K. Togano, H. Kito, H. Ihara
and S. Shin: Phys. Rev. Lett. {\bf 87} (2001) 177006.
\bibitem{sza01}P. Szabo, P. Samuely, J. Kacmarcik, T. Klein, J. Marcus, D. Fruchart,
S. Miraglia, C. Marcenat and A. G. M. Jansen: Phys. Rev. Lett. {\bf 87} (2001) 137005.
\bibitem{che01}X. K. Chen, M. J. Konstantinovic, J. C. Irwin, D. D. Lawrie and 
J. P. Frank: Phys. Rev. Lett. {\bf 87} (2001) 157002.
\bibitem{giu01}F. Giubileo, D. Roditchev, W. Sacks, R. Lamy, D.X. Thanh, J. Klein,
S. Miraglia, D. Fruchart, J. Marcus and Ph. Monod: Phys. Rev. Lett. {\bf 87}
(2001) 177008.
\bibitem{bou01}F. Bouquet, R. A. Fisher, N. E. Philips, D. G. Hinks and 
J. D. Jorgensen: 
Phys. Rev. Lett. {\bf 87} (2001) 047001.
\bibitem{lim01}O. F. de Lima, C. A. Cardoso, R. A. Ribeiro, M. A. Avilla and
A. A. Coelho: Phys. Rev. B{\bf 64} (2001) 144517.
\bibitem{xu01}M. Xu, H. Kitazawa, Y. Takano, J. Ye, K. Nishida, H. Abe, 
A. Matsushita, N. Tsuji and G. Kido: Appl. Phys. Lett. {\bf 79} (2001) 2779.
\bibitem{ang02}M. Angst, R. Puzniak, A. Wisniewski, J. Jun, S. M. Kazakov, 
J. Karpinski, J. Roos and H. Keller: Phys. Rev. Lett. {\bf 88} (2002) 167004.
\bibitem{kor01}J. Kortus, I. I. Mazin, K. D. Belashchenko, V. P. Antropov and
L. L. Boyer: Phys. Rev. Lett. {\bf 86} (2001) 4656.
\bibitem{uch02}H. Uchiyama, K. M. Shen, S. Lee, A. Damascelli, D. H. Lu, D. L. Feng,
Z. X. Shen, and S. Tajima: Phys. Rev. Lett. {\bf 88} (2002) 157002.
\bibitem{haa01}S. Haas and K. Maki: Phys. Rev. B{\bf 65} (2001) 020502.
\bibitem{mir02}P. Miranovic, K. Machida and V.G. Kogan: J. Phys. Soc. Jpn. {\bf 72}
(2003) 221.
\bibitem{nak02}N. Nakai, M. Ichioka and K. Machida: J. Phys. Soc. Jpn. {\bf 71}
(2002) 23.
\bibitem{tew02}L. Tewordt and D. Fay: Phys. Rev. Lett. {\bf 89} (2002) 137003.
\bibitem{bou02}F. Bouquet, Y. Wang, I. Sheikin, T. Plackowski, A. Junod, S. Lee
and S. Tajima: Phys. Rev. Lett. {\bf 89} (2002) 257001.
\bibitem{mat58}D. C. Mattis and J. Bardeen: Phys. Rev. {\bf 111}, 412 (1958).
\bibitem{kai02}R. A. Kaindl, M. A. Carnahan, J. Orenstein and D. S. Chemla: 
Phys. Rev. Lett. {\bf 88} (2002) 027003.
\bibitem{yana01}T. Yanagisawa, S. Koikegami, H. Shibata, S. Kimura, S. Kashiwaya,
A. Sawa, N. Matsubara and K. Takita: J. Phys. Soc. Jpn. {\bf 70} (2001) 2833.
\bibitem{glo57}R. E. Glover and M. Tinkham: Phys. Rev. {\bf 108} (1957) 243.
\bibitem{bel01}K. D. Belashchenko, M. van Schilfgaarde and V. P. Antropov: 
Phys. Rev. B{\bf 64} (2001) 092503.



\end{thebibliography}
\end{document}